\documentstyle[a4,11pt]{article}

\begin{document}

\begin{titlepage}
\vskip 2cm
\begin{flushright}
Preprint CNLP-1995-08
\end{flushright}
\vskip 2cm
\begin{center}
{\large {\bf On some soliton equations in 2+1 dimensions and their
1+1 and/or 2+0 dimensional integrable reductions}}\footnote{Preprint
CNLP-1995-08.Alma-Ata.1995 \\
cnlpgnn@satsun.sci.kz}
\vskip 2cm

{\bf F.B.Altynbaeva, A.K.Danlybaeva, G.N.Nugmanova and R.N.Syzdykova }

\end{center}
\vskip 1cm

 Centre for Nonlinear Problems, PO Box 30, 480035, Almaty-35, Kazakstan

\begin{abstract}
Some soliton equations in 2+1 dimensions and their 1+1 and/or 2+0 dimensional
integrable reductions are considered.
\end{abstract}


\end{titlepage}

\setcounter{page}{1}
\newpage

\tableofcontents
\section{Introduction}

Consider the Myrzakulov IX (M-IX) equation
$$ iS_t+\frac{1}{2}[S,M_1S]+A_2S_x+A_1S_y = 0 \eqno(1a)$$
$$ M_2u=\frac{\alpha^{2}}{2i}tr(S[S_x,S_y]) \eqno(1b)$$
where $ \alpha,b,a  $=  consts and
$$
S= \pmatrix{
S_3 & rS^- \cr
rS^+ & -S_3
},\quad S^{\pm}=S_{1}\pm iS_{2} \quad  S^2 = EI,\quad r^{2}=\pm 1,
$$
$$
M_1= \alpha ^2\frac{\partial ^2}{\partial y^2}+4\alpha (b-a)\frac{\partial^2}
   {\partial x \partial y}+4(a^2-2ab-b)\frac{\partial^2}{\partial x^2},
$$
$$
M_2=\alpha^2\frac{\partial^2}{\partial y^2} -2\alpha(2a+1)\frac{\partial^2}
   {\partial x \partial y}+4a(a+1)\frac{\partial^2}{\partial x^2},
$$
$$
A_1=i\{\alpha (2b+1)u_y - 2(2ab+a+b)u_{x}\},
$$
$$
A_2=i\{4\alpha^{-1}(2a^2b+a^2+2ab+b)u_x - 2(2ab+a+b)u_{y}\}.
$$

This equation was introduced in [1] and  arises from the compatibility
condition of the following linear equations
$$ \alpha \Phi_y =\frac{1}{2}[S+(2a+1)I]\Phi_x \eqno(2a) $$
$$ \Phi_t=2i[S+(2b+1)I]\Phi_{xx}+W\Phi_x \eqno(2b) $$
with
$$ W = 2i\{(2b+1)(F^{+} + F^{-} S) +(F^{+}S + F^{-}) +
(2b-a+\frac{1}{2})SS_x+\frac{1}{2}S_{x} + \frac{\alpha}{2} SS_y \}, \quad
$$
$$
F^{\pm} = A \pm D , \quad A=i[u_{y} - \frac{2a}{\alpha} u_{x}],
\quad D=i[\frac{2(a+1)}{\alpha} u_{x} - u_{y}].
$$

It is well known that equation (1) is gauge and Lakshmanan equivalent
to the following Zakharov equation(ZE)[2]
$$
iq_{t}+M_{1}q+vq=0, \eqno(3a)
$$
$$
ip_{t}-M_{1}p-vp=0, \eqno(3b)
$$
$$
M_{2}v = -2M_{1}(pq), \eqno(3c)
$$

The Lax representation of this equation has the form[2]
$$
\alpha \Psi_y =B_{1}\Psi_x + B_{0}\Psi, \eqno(4a)
$$
$$
\Psi_t=iC_{2}\Psi_{xx}+C_{1}\Psi_x+C_{0}\Psi, \eqno(4b)
$$
with
$$
B_{1}= \pmatrix{
a+1 & 0 \cr
0   & a
},\quad
B_{0}= \pmatrix{
0   &  q \cr
p   &  0
}
$$
$$
C_{2}= \pmatrix{
b+1 & 0 \cr
0   & b
},\quad
C_{1}= \pmatrix{
0   &  iq \cr
ip  &  0
},\quad
C_{0}= \pmatrix{
c_{11}  &  c_{12} \cr
c_{21}  &  c_{22}
}
$$
$$
c_{12}=i(2b-a+1)q_{x}+i\alpha q_{y},\quad
c_{21}=i(a-2b)p_{x}-i\alpha p_{y}.
$$
Here $c_{jj}$ is the solution of the  following  equations
$$
(a+1) c_{11x}- \alpha c_{11y} = i[(2b-a+1)(pq)_{x} + \alpha (pq)_{y}],
$$
$$
ac_{22x}-\alpha c_{22y} = i[(a-2b)(pq)_{x} - \alpha (pq)_{y}]. \eqno (5)
$$

Below we consider the  some reductions of the M-IX equation(1) and their
gauge equivalent counterparts.

\section{Some $\sigma$-models with potentials
and  the Klein-Gordon-type equations}

First, let us consider   the some  $\sigma$-models with potentials, which
are  the some reductions of equation(1) in 1+1 or 2+0 dimensions.

\subsection{The Myrzakulov XXXII (M-XXXII) equation}

Suppose that now $\nu = t$ is the some "hidden" parameter, $S = S(x,y), \quad
u = u(x,y), \quad  q = q(x,y), \quad p = p(x,y), \quad v = v(x,y) $ and
at the same time $\Phi = \Phi(x,y,\nu), \quad
\Psi = \Psi(x,y,\nu)$.
Then equation (1) reduces to the following $\sigma$-model with potential
$$
M_1S + \{ k_{1} S^{2}_{x} + k_{2} S_{x}S_{y} + k_{3} S^{2}_{y} \} S
+ A_2 SS_x+A_1 SS_y = 0 \eqno(6a)$$
$$ M_2u=\frac{\alpha^{2}}{2i}tr(S[S_x,S_y]) \eqno(6b)$$
where $M_{1}$ we write in the form
$$
M_1= k_{3}\frac{\partial ^2}{\partial y^2}+k_{2}\frac{\partial^2}
   {\partial x \partial y}+k_{1}\frac{\partial^2}{\partial x^2},
$$
which is the compatibility condition $ \Phi_{y\nu} = \Phi_{\nu y}$
of the set (2) and called the M-XXXII equation[1]. The G-equivalent
and L-equivalent counterpartt of this equation is given by
$$
M_{1}q+vq=0, \quad M_{1}p + vp=0, \quad M_{2}v = -2M_{1}(pq). \eqno(7)
$$
This is the some modified complex Klein-Gordon equation (mKGE).

\subsection{The Myrzakulov XV (M-XV) equation}

Let us consider the case: $a = b$. Then we obtain the M-XV equation [1]
$$
\alpha^{2} S_{yy} - a(a+1) S_{xx} +
\{\alpha^{2} S^{2}_{y} - a(a+1) S^{2}_{x}\} S +
+  A_{2}^{\prime \prime } SS_x + A_{1}^{\prime \prime } SS_y  \eqno(8a)
$$
$$
M_{2} u = \frac{\alpha^{2}}{2i} tr( S [S_{x}, S_{y}]) \eqno(8b)
$$
where $A_{j}^{\prime \prime} = A_{j}$ as $ a = b$. The corressponding mKGE
has the form
$$
\alpha^{2}q_{yy} - a(a+1)q_{xx} + v q = 0, \quad
M_{2} v = -2[\alpha^{2} (\mid  q \mid^{2})_{yy} -
a(a+1) (\mid q\mid^{2})_{xx}] \eqno(9)
$$
The Lax representations of these equations we obtain from (2) and (4)
respectively as $a = b$.

\subsection{The Myrzakulov XIV (M-XIV) equation}

Now we  consider the reduction: $a=-\frac{1}{2}$. Then the equation (1)
reduces to the M-XIV equation[1]
$$
S_{xx} + 2\alpha(2b+1)S_{xy}+\alpha^{2}S_{yy}) +
\{S_{xx} + 2\alpha(2b+1)S_{xy}+\alpha^{2}S_{yy}\}S +
A^{\prime}_{2}SS_x+A^{\prime}_{1}SS_y = 0 \eqno(10a)
$$
$$
\alpha^{2}u_{yy} - u_{xx}=
\frac{\alpha^{2}}{2i}tr(S[S_x,S_y]) \eqno(10b)
$$
where $A^{\prime}_{j} = A_{j}$ as $a=-\frac{1}{2}$. The corresponding gauge
equivalent equation is obtained from (3) and looks like
$$
q_{xx} + 2\alpha(2b+1)q_{xy}+\alpha^{2}q_{yy}+vq = 0 \eqno(11a)
$$
$$
\alpha^{2}v_{yy} - v_{xx}=-2\{\alpha^{2}(pq)_{yy} +
2\alpha (2b+1)(pq)_{xy} +(pq)_{xx}\} \eqno(11b)
$$
From (2) and (4) we obtain the Lax representations of (10) and (11)
respectively as $ a = -\frac{1}{2}$.

\subsection{The Myrzakulov XIII (M-XIII) equation}

Now let us consider the case: $ a=b=-\frac{1}{2} $. In this case the
equations (1) reduce to the $\sigma$-model
$$
S_{xx}+\alpha^{2}S_{yy} +
\{S^{2}_{x}+\alpha^{2}S^{2}_{y}\}S +
iu_{y}SS_x+iu_{x}SS_y = 0 \eqno(12a)
$$
$$
\alpha^{2}u_{yy} - u_{xx} = \frac{\alpha^{2}}{2i}tr(S[S_y,S_x]) \eqno(12b)
$$
which is the M-XIII equation[1].
The  equivalent counterpart of the equation(12) is the following  equation
$$
q_{xx}+\alpha^{2}q_{yy}+ vq = 0 \eqno(13a)
$$
$$
\alpha^{2}v_{yy} - v_{xx}=-2\{\alpha^{2}(pq)_{yy}
 +(pq)_{xx}\} \eqno(13b)
$$
that follows from the mKGE(7).

\subsection{The Myrzakulov XII (M-XII) equation}

Now let us consider the case: $ a=b=-1$. In this case the
equations(6) reduce to the $\sigma$-model - the M-XII equation[1]
$$
S_{YY} +  S^{2}_{Y} S + iw SS_{Y} = 0 \eqno(14a)
$$
$$
w_{Y} + w_{X} +\frac{1}{4i}tr(S[S_X,S_Y]) = 0 \eqno(14b)
$$
where $X=2x, \quad Y = \alpha y, \quad w = -\frac{u_{Y}}{\alpha}$.
The  equivalent counterpart of the equation(14) is the mKGE
$$
q_{YY}+ vq = 0 \eqno(15a)
$$
$$
v_{X} + v_{Y} + 2(pq)_{Y} = 0. \eqno(15b)
$$
that follows from the (7).

\subsection{The Myrzakulov XXXI (M-XXXI) equation}

This $\sigma$ -model equation is read as[1]
$$
bS_{\eta \eta} - (b+1) S_{\xi \xi}] +
\{bS_{\eta \eta} - (b+1) S_{\xi\xi}\} S +
i(b+1)w_{\eta} SS_{\eta} + ibw_{\xi}SS_{\xi} =0 \eqno(16a)
$$
$$
w_{\xi \eta} = - \frac{1}{4i}tr(S[S_{\eta},S_{\xi}]) \eqno(16b)
$$
which is the M-XXXI equation, where $ w = -\alpha^{-1} u$.
The   equivalent mKGE looks like
$$
(1 + b)q_{\xi \xi } - b q_{\eta \eta } + vq = 0 \eqno(17a)
$$
$$
v_{\xi \eta } = -2\{(1+ b) (pq)_{\xi \xi} - b(pq)_{\eta \eta}\} \eqno(17b)
$$

So we have found the G-equivalent counterparts of
the $\sigma$-models with potentials.

\section{The (2+1)-dimensional integrable spin systems and NLSE type
equations}

The equation (1) admits  some interesting (2+1)-dimensional
integrable reductions. Let us now consider these particular integrable
cases.

\subsection{The M-VIII equation}

Let $b=0$. Then the equations (1) take the form
$$
iS_t=\frac{1}{2}[S_{\xi\xi},S]+iwS_{\xi}   \eqno (18a)
$$
$$
w_{\eta}= \frac{1}{4i}tr(S[S_{\eta},S_{\xi}])  \eqno (18b)
$$
where
$$
\xi = x+\frac{a+1}{\alpha}y,\quad
\eta = -x -\frac{a}{\alpha}y,\quad w = - \frac{1}{\alpha}u_{\xi}
$$
which is the M-VIII equation[1]. The corresponding Lax representation
has the form
$$
\Phi_{Z^{+}} = S\Phi_{Z^{-}} \eqno(19a)
$$
$$
\Phi_{t} = 2i[S+I]\Phi_{Z^{-}Z^{-}} + i K\Phi_{Z^{-}} \eqno(19b)
$$
where  $Z^{\pm} = \xi \pm \eta$ and
$$
K = E_{0} + E_{0}S + S_{Z^{-}} + 2SS_{Z^{-}} + SS_{Z^{+}}, \quad
E_{0} = \frac{i}{\alpha}(u_{Z^{+}} + u_{Z^{-}}).
$$

The gauge equivalent counterpart
of the  equation(18) we obtain from(3) as $b=0$
$$
iq_{t}+q_{\xi \xi}+vq=0, \eqno(20a)
$$
$$
v_{\eta} = -2r^{2}(\bar q q)_{\xi}, \eqno(20b)
$$
which is the other Zakharov equation[2].

\subsection{The Ishimori equation}

Now let us consider the case: $ a=b=-\frac{1}{2} $. In this case the
equations(1) reduce to the well-known Ishimori equation
$$
iS_t+\frac{1}{2}[S,(\frac{1}{4}S_{xx}+\alpha^{2}S_{yy})]+
iu_{y}S_x+iu_{x}S_y = 0 \eqno(21a)
$$
$$
\alpha^{2}u_{yy} - \frac{1}{4}u_{xx}=
\frac{\alpha^{2}}{4i}tr(S[S_y,S_x]). \eqno(21b)
$$

The gauge equivalent counterpart of the equation(21) is the Davey-Stewartson
equation
$$
iq_t+\frac{1}{4}q_{xx}+\alpha^{2}q_{yy}+
vq = 0 \eqno(22a)
$$
$$
\alpha^{2}v_{yy} - \frac{1}{4}v_{xx}=-2\{\alpha^{2}(pq)_{yy}
 +\frac{1}{4}(pq)_{xx}\} \eqno(22b)
$$
that follows from the ZE(3). This fact was for first time
established in [3].  The Lax representations of (21) and (22)
we can get from
(2) and (4)
respectively as $ a = b = - \frac{1}{2}$.

\subsection{The M-XVIII equation}

Now we  consider the reduction: $a=-\frac{1}{2}$. Then the equation (1)
reduces to the M-XVIII equation[1]
$$
iS_t+\frac{1}{2}[S,(\frac{1}{4}S_{xx}-\alpha(2b+1)S_{xy}+\alpha^{2}S_{yy})]+
A^{\prime}_{2}S_x+A^{\prime}_{1}S_y = 0 \eqno(23a)
$$
$$
\alpha^{2}u_{yy} - \frac{1}{4}u_{xx}=\frac{\alpha^{2}}{4i}tr(S[S_y,S_x]) \eqno(23b)
$$
where $A^{\prime}_{j} = A_{j}$ as $a=-\frac{1}{2}$. The corresponding gauge
equivalent equation is obtained from(3) and looks like
$$
iq_t+\frac{1}{4}q_{xx}-\alpha(2b+1)q_{xy}+\alpha^{2}q_{yy}+
vq = 0 \eqno(24a)
$$
$$
\alpha^{2}v_{yy} - \frac{1}{4}v_{xx}=-2\{\alpha^{2}(pq)_{yy}-
\alpha (2b+1)(pq)_{xy} +\frac{1}{4}(pq)_{xx}\} \eqno(24b)
$$
From (2) and (4) we obtain the Lax representations of (23) and (24)
respectively as $ a = -\frac{1}{2}$.

\subsection{The M-XIX equation}

Let us consider the case: $a = b$. Then we obtain the M-XIX equation [1]
$$
iS_t = \frac{1}{2} [S,  \{\alpha^{2} S_{yy} - a(a+1) S_{xx}\}]
+  A_{2}^{\prime \prime } S_x + A_{1}^{\prime \prime }  S_y  \eqno(25a)
$$
$$
M_{2} u = -\frac{\alpha^{2}}{4i} tr( S [S_{x}, S_{y}]) \eqno(25b)
$$
where $A_{j}^{\prime \prime} = A_{j}$ as $ a = b$. The corressponding NLSE
has the form
$$
iq_t +  \alpha^{2}q_{yy} - a(a+1)q_{xx} + v q = 0, \eqno(26a)
$$
$$
M_{2} v = -2[\alpha^{2} (\mid  q \mid^{2})_{yy} -
a(a+1) (\mid q\mid^{2})_{xx}] \eqno(26b)
$$
The Lax representations of these equations we obtain from (2) and (4)
respectively as $a = b$.

\subsection{The M-XX equation}

This equation is read as[1]
$$
iS_t = \frac{1}{2}[S,bS_{\eta \eta} - (b+1) S_{\xi \xi}]+
w_{\eta} S_{\eta} + w_{\xi}S_{\xi} \eqno(27a)
$$
$$
w_{\xi \eta} = - \frac{1}{4i}tr(S[S_{\eta},S_{\xi}]) \eqno(27b)
$$
where $ w = -\alpha^{-1} u$.
The associated linear problem is given by
$$
\Phi_{Z^{+}} = S\Phi_{Z^{-}} \eqno(28a)
$$
$$
\Phi_{t} = 2i[S+(2b+1)I]\Phi_{Z^{-}Z^{-}} + i W_{0}\Phi_{Z^{-}} \eqno(28b)
$$
where  $Z^{\pm} = \xi \pm \eta$ and
$$
W_{0} = (2b+1)(E+FS) + F + ES + \frac{1}{2} S_{Z^{-}} + 2(2b + 1) SS_{Z^{-}}
+ SS_{Z^{+}}
$$
$$
E = \frac{i}{\alpha}u_{Z^{-}}, \quad F = \frac{i}{\alpha}u_{Z^{+}}.
$$

The  gauge equivalent equation looks like
$$
iq_t+ (1 + b)q_{\xi \xi } - b q_{\eta \eta } + vq = 0 \eqno(29a)
$$
$$
v_{\xi \eta } = -2\{(1+ b) (pq)_{\xi \xi} - b(pq)_{\eta \eta}\} \eqno(29b)
$$

This equation is integrated by the linear problem
$$
f_{Z^{+}} = \sigma_{3} f_{Z^{-}} + B_{0}f \eqno(30a)
$$
$$
f_{t} = 4i C_{2} f_{Z^{-}Z^{-}} + 2 C_{1} f_{Z^{-}} + C_{0}f. \eqno(30b)
$$
where $B_{0},\quad C_{j}$ are given as in  (4).

Thus, we have presented the some reductions of the equation (1). All of
these reductions are integrable in the sense that they admit the Lax
representations.

\section{The (2+1)-dimensional integrable spin system with anisotropy}

\subsection{Gauge equivalent counterpart of the anisotropic spin system}

As integrable one, the anisotropic LLE can admits the several integrable
(2+1)-dimensional extensions[1].
One of the such integrable (2+1)-dimensional extension of the LLE
as $ J_{1} = J_{2} = 0, J_{3} = \triangle$ is the
following M-I equation with one-ion anisotropy
$$
{\bf S}_t = ({\bf S} \wedge {\bf S}_{y} + u {\bf S})_{x} +
v {\bf S} \wedge {\bf n}, \eqno(31a)
$$
$$
u_{x} = - {\bf S} \cdot ({\bf S}_{x} \wedge {\bf S}_{y}),  \eqno(31b)
$$
$$
v_x =
\triangle ({\bf S}_y \cdot {\bf n}) \eqno(31c)
$$
where $ u $ and $ v $ are scalar functions, $ {\bf n}=(0,0,1)$, and
$\triangle<0$ and $\triangle>0 $ correspond respectively to the system with
an easy plane and to that with an easy axis. Note that if the symmetry
$ \partial_x=\partial_y $ is imposed then the M-I equation (31) reduces
to the well-known LLE with single-site anisotropy
$$
{\bf S}_t = {\bf S} \wedge ({\bf S}_{xx} +
\triangle ({\bf S} \cdot {\bf n}){\bf n}) \eqno(32)
$$
which is the particular case of the anisotropic LLE as
$J_{1}=J_{2}=0, J_{3}=\triangle$.  On the other hand, in the
case when $\triangle = 0$, the
equation(31) becomes the isotropic M-I equation.
It is known that the equation(32) is gauge equivalent to the NLSE [4-7].
In this subsection we construct the NLSE which is gauge equivalent to the
equation(31) with the easy-axis
anisotropy $ (\triangle > 0). $

The Lax representation of the  equation (31) may be given by[6]
$$
\psi_x=L_1 \psi,\quad \psi_t=2\lambda \psi_y+M_1 \psi \eqno (33)
$$
where
$$
L_1=i \lambda S+ \mu[\sigma_3, S],\quad
M_1=2\lambda A+2i\mu[A,\sigma_3]+4i \mu^2 \{\sigma_3, V\}\sigma_3 \eqno(34)
$$
with
$$
S=\sum_{k=1}^3 S_k \sigma_k,\quad
A=\frac{1}{4}([S,S_y]+2iuS), \quad \mu=\sqrt\frac{\triangle}{4}, \quad \triangle >0.
$$
and
$$
V=\triangle\int^{x}_{-\infty} S_y dx. \eqno(35)
$$
Here $ \sigma_k $ is Pauli matrices, [,] (\{,\}) denoting commutator
(anticommutator),
and $\lambda $ is a spectral parameter. The matrix $ S $ has the following
properties: $ S^2=I, \quad S^{\ast} = S,\quad trS=0 $.
The compatibility condition of system (33) $ \psi_{xt}=\psi_{tx} $ gives the
 equation (31).
Let us now consider the gauge transformation induced by $ g(x,y,t): \psi=
g^{-1}\phi$, where $g^{\ast}=g^{-1}\in SU(2)$. It follows from the properties
of the matrix S that it can be represented in the form $ S=
g^{-1}\sigma_3g$. The new
gauge equivalent operators $ L_2, M_2 $ are given by
$$
L_2=gL_1g^{-1}+g_xg^{-1},\quad M_2=gM_1g^{-1}+g_tg^{-1}
- 2\lambda g_{y}g^{-1}  \eqno(36)
$$
and satisfy the following system of equations
$$
\phi_x=L_2\phi,\quad \phi_t=2\lambda\phi_y+M_2\phi. \eqno(37)
$$
Now choosing
$$
g_xg^{-1} + \mu g[\sigma_{3},S]g^{-1}=U_{0},\,\,\,gSg^{-1}=\sigma_{3}
 \eqno(38a)
$$
$$
g_t g^{-1}+2i\mu g[A,\sigma_3]g^{-1}+
4i\mu^2 g \{\sigma_3,V\}\sigma_3g^{-1}=V_{0} \eqno(38b)
$$
with
$$
U_{0} =
\left ( \begin{array}{cc}
0       & q \\
-\bar q & 0
\end{array} \right),\,\,\,\,\,
V_{0} =
i\sigma_{3}(\partial^{-1}_{x}|q|^{2}_{y} - U_{0y})
$$
where $ q(x,y,t) $ the new complex valued fields. Hence we finally obtain
$$
L_2=i\lambda \sigma_3+U_{0},\,\,\,\,\, M_2=V_{0} \eqno(39)
$$
The compatibility condition $ \phi_{xt}=\phi_{tx} $ of the system (37) with
the operators $ L_2, M_2$ (39) leads to the (2+1)-dimensional NLSE[2,8]
$$
iq_t=q_{xy}+wq,\quad w_x =2(|q|^2)_y.  \eqno(40)
$$

This equation under the reduction $ \partial_{y} = \partial_{x} $
equation(40) becomes the well known
(1+1)-dimensional NLSE. Thus we have shown that the M-I
equation with
single-site anisotropy(31)  is gauge equivalent to the (2+1)-dimensional
NLSE - the
Zakharov equation(40).

\subsection{The isotropic and anisotropic spin systems: gauge equivalence}

It is  known  that the equations (40) is gauge and
geometrical equivalent to the isotropic M-I equation
$$
iS^{\prime}_t=\frac{1}{2}([S^{\prime}, S^{\prime}_y]+2iu^{\prime}
S^{\prime})_x \eqno(41a)
$$
$$
u^{\prime}_x = - \frac{1}{4i}tr (S^{\prime}[S^{\prime}_{x}, S^{\prime}_{y}]) \eqno(41b)
$$
which was introduced in[1] and arises from the compatibility condition
of the linear problem
$$
f_x = L^{\prime}_{1} f,\quad f_{t} =2\lambda f_y+
\lambda M^{\prime}_{1}f \eqno(42)
$$
where
$$
L^{\prime}_1=i\lambda S^{\prime},\quad  M^{\prime}_1=\frac{1}{2}
([S^{\prime},S^{\prime}_y]+2iu^{\prime}S^{\prime}). \eqno(43)
$$

Now we show that  between the isotropic(41) and anisotropic(31) versions
of the M-I  equation  the gauge equivalence takes place.
Indeed the Lax representations, (33) and (42),
which reproduce equations (31) and (41), respectively can be obtained from
each other by the $ \lambda $-independent gauge transformation
$$ L_{1}^{\prime}=hL_{1}h^{-1}+h_{x}h^{-1},\quad
M^{\prime}_{1}=hM_{1} h^{-1}+h_{t} h^{-1}. \eqno(44)
$$
with $ h(x,y,t)=
\psi^{-1}|_{\lambda=\lambda_{0}} $, where $ \lambda_{0}$ is some fixed
value of the spectral parameter $\lambda$.

So, the solutions of equations(31) and (41) are connected with each other by formulas
$ S=h^{-1} S^{\prime}h. $ Now we present the important relations between the
field
variables $ q $ and $S$:
$$
\mid q \mid^2 = \frac{1}{2}[{\bf S}^{2}_{x}-8\mu S_{3x}+16\mu^{2}(1-S^{3}_{2})]
\eqno(45a)
$$
$$
\bar q_{x} q  - \bar q q_{x} =
\frac{i}{4}{\bf S}\cdot{[{\bf S_{xx}}+16\mu^2({\bf S}\cdot
{\bf n}){\bf n})]+4\mu{\bf S}\cdot({\bf S}_{xx}\wedge {\bf n})} \eqno(45b)
$$

These relations coincide with the corresponding connections between
$ q $ and $S$ from the one-dimensional case.

Note that the equation(31)  with ($\triangle < 0$)
easy
plane single-site anisotropy is gauge equivalent to the following  general (2+1)-
dimensional NLSE[2,8]
$$
iq_t=q_{xy}+wq \eqno(46a)
$$
$$
ip_t=-p_{xy}-wp \eqno(46b)
$$
$$
w_x =2(pq)_y. \eqno(46c)
$$
Besides, there  can be shown that the equation(31), when
$ S \in SU(1,1)/U(1)$, i.e. the non-compact group case, is gauge
equivalent to the NLSE(46) with the repulsive interaction,
$p=-\bar q$.

Finally, we note that the M-I equation(31) is the particular case of the
M-III equation
$$
{\bf S}_{t}=({\bf S}\wedge {\bf S}_{y}+u{\bf S})_x+2b(cb+d){\bf S}_{y}
     -4cv{\bf S}_{x} + {\bf S} \wedge {\bf V} \eqno (47a)
$$
$$
u_x=-{\bf S}\dot({\bf S}_{x}\wedge {\bf S}_{y}), \eqno(47b)
$$
$$
v_x=\frac{1}{4(2bc+d)^2}
   ({\bf S}^{2}_{x})_y \eqno (47c)
$$
$$
{\bf V}_x= J{\bf S}_{y}. \eqno(47d)
$$
Note these equations admit the some integrable reductions:
a) the isotropic M-I equation as $ c = J = 0$;
b) the anisotropic M-I equation as $c = J_{1} = J_{2} = 0;$
c) the M-II equation  as $ d = J_{j} =0$;
d) the isotropic M-III equation as $ J = 0$ and so on[1].

\section{Integrable spin-phonon systems and the Yajima-Oikawa and Ma
equations}

Let us now we consider the reduction of the M-IX equation (1) as $ a=b=-1$.
We have
$$
iS_t+\frac{1}{2}[S,S_{YY}]+iwS_Y = 0 \eqno(48a)
$$
$$
w_{X} + w_{Y} + \frac{1}{4i}tr(S[S_X,S_Y]) = 0 \eqno(48b)
$$
This equation is the Myrzakulov VIII (M-VIII) equation[1].  The G-equivalent
and L-equivalent counterpart of equation (48) is given by
$$
iq_{t}+ q_{YY} + vq = 0, \eqno(49a)
$$
$$
ip_{t}-p_{YY} - vp = 0, \eqno(49b)
$$
$$
v_{X} + v_{Y} + 2(pq)_{Y} =0.  \eqno(49c)
$$

Now let us take the case when $ X = t$. Then the M-VIII equation (48) pass
to the following M-XXXIV equation
$$
iS_t+\frac{1}{2}[S,S_{YY}]+iwS_Y = 0 \eqno(50a)
$$
$$
w_{t} + w_{Y} + \frac{1}{4}\{tr(S^{2}_{Y})\}_{Y} = 0 \eqno(50b)
$$

The  M-XXXIV  equation (50)
was proposed in [1] to describe a nonlinear dynamics of the
compressible magnets (see, Appendix). It is integrable and has  the different soliton
solutions[1].

In our case  equation (49)  becomes
$$
iq_{t}+ q_{YY} + vq = 0, \eqno(51a)
$$
$$
ip_{t}-p_{YY} - vp = 0, \eqno(51b)
$$
$$
v_{t} + v_{Y} + 2(pq)_{Y} =0.  \eqno(51c)
$$
that is the Yajima-Oikawa equation(YOE)[9]. So we have proved that the
M-XXXIV equation (50) and the YOE (51) is gauge equivalent
to each other.  The Lax representations of (50) and (51) we can get from
(2) and (4)
respectively as $ a = b = - 1$( see, for example, the ref.[1]). Note that
our Lax representation for the YOE (51) is different than that which was
presented in [9].

Also we would like note that the M-VIII equation (48) we usually write in
the following form
$$
iS_t=\frac{1}{2}[S_{\xi\xi},S]+iwS_{\xi}   \eqno (52a)
$$
$$
w_{\eta}= \frac{1}{4i}tr(S[S_{\eta},S_{\xi}])  \eqno (52b)
$$

The gauge equivalent counterpart
of this  equation is the following ZE[2]
$$
iq_{t}+q_{\xi \xi}+vq=0, \eqno(53a)
$$
$$
v_{\eta} = -2r^{2}(\bar q q)_{\xi}. \eqno(53b)
$$

As $ \eta = t$  equation (52) take the other form of the M-XXXIV equation
$$
iS_t=\frac{1}{2}[S_{\xi\xi},S]+iwS_{\xi}   \eqno (54a)
$$
$$
w_{t}= \frac{1}{4i}\{tr(S^{2}_{\xi})\}_{\xi}  \eqno (54b)
$$

Similarly, equation (53) becomes
$$
iq_{t}+q_{\xi \xi}+vq=0, \eqno(55a)
$$
$$
v_{t} = -2r^{2}(\bar q q)_{\xi}, \eqno(55b)
$$

which is called the Ma equation and  was considered in [10].

\section{Conclusion}

The some integrable reductions  of the Myrzakulov IX equation are considered.
We have established the gauge equivalence between the (1+1)-, or (2+0)-
dimensional $\sigma$-models and the Klein-Gordon type equations.
Also we have shown that  the Myrzakulov XXXIV equation
which describe nonlinear dynamics of compressible magnets and
the Yajima-Oikawa-Ma equations are gauge equivalent to each other[11-13].

Finally, we note that between the some above mentioned equations take place
the Lakshmanan equivalence (see, for example, the refs.[14-16]).

\section{Appendix: On some soliton equations of compressible magnets}

Solitons in magnetically ordered crystals have been widely investigated
from both theoretical and experimental points of view. In particular, the
existence of coupled magnetoelastic solitons in the Heisenberg compressible
spin chain has been extensively demonstrated. In [1]
were presented a new classes of integrable and nonintegrable soliton
equations of spin systems. Below we present the some of these  nonlinear
models of magnets - the some of the Myrzakulov equations(ME), which
describe the nonlinear dynamics of compressible magnets.

\subsection{The 0-class of spin-phonon systems}

The Myrzakulov equations with the potentials have the form:\\
the Myrzakulov LVII(M-LVII) [the $ M^{10}_{00}$ - equation]:
$$ 2iS_t=[S,S_{xx}]+(u+h)[S,\sigma_3]  $$
the Myrzakulov LVI(M-LVI) [the $ M^{20}_{00}$ - equation]:
$$ 2iS_t=[S,S_{xx}]+(uS_3+h)[S,\sigma_3]  $$
the Myrzakulov LV(M-LV) [the $ M^{30}_{00}$ - equation]:
$$ 2iS_t=\{(\mu \vec S^2_x-u+m)[S,S_x]\}_x+h[S,\sigma_3]  $$
the Myrzakulov LIV(M-LIV) [the $ M^{40}_{00}$ - equation]:
$$ 2iS_t=n[S,S_{xxxx}]+2\{(\mu \vec S^2_x-u+m)[S,S_x]\}_x+
h[S,\sigma_3] $$
the Myrzakulov LIII(M-LIII) [the $ M^{50}_{00}$ - equation]:
$$ 2iS_t=[S,S_{xx}]+2iuS_x  $$
where $v_{0}, \mu, \lambda, n, m, a, b, \alpha, \beta, \rho, h$ are constants,
$u$ is a scalar function(potential),
subscripts denote partial differentiations, $[,]$ (\{,\}) is
commutator (anticommutator),
$$S= \pmatrix{
S_3 & rS^- \cr
rS^+ & -S_3
}, \,\,\,\,\, S^{\pm}=S_{1}\pm i S_{2},\,\,\,\, r^{2}=\pm 1\,\,\,\,\, S^2=I.    $$

\subsection{The 1-class of spin-phonon systems}

The Myrzakulov LII(M-LII) [the $ M^{11}_{00}$ - equation]:
$$
2iS_t=[S,S_{xx}]+(u+h)[S,\sigma_3]
$$
$$
\rho u_{tt}=\nu^2_0 u_{xx}+\lambda(S_3)_{xx}
$$
the Myrzakulov LI(M-LI) [the $ M^{12}_{00}$ - equation]:
$$
2iS_t=[S,S_{xx}]+(u+h)[S,\sigma_3]
$$
$$
\rho u_{tt}=\nu^2_0 u_{xx}+\alpha(u^2)_{xx}+\beta u_{xxxx}+
    \lambda(S_3)_{xx}
$$
the Myrzakulov L(M-L) [the $ M^{13}_{00}$ - equation]:
$$
2iS_t=[S,S_{xx}]+(u+h)[S,\sigma_3]
$$
$$
u_t+u_x+\lambda(S_3)_x=0
$$
the Myrzakulov XXXXIX(M-XXXXIX) [the $ M^{14}_{00}$ - equation]:
$$
2iS_t=[S,S_{xx}]+(u+h)[S,\sigma_3]
$$
$$
u_t+u_x+\alpha(u^2)_x+\beta u_{xxx}+\lambda(S_3)_x=0
$$

\subsection{The 2-class of spin-phonon systems}

The Myrzakulov XXXXVIII(M-XXXXVIII) [the $ M^{21}_{00}$ - equation]:
$$
2iS_t=[S,S_{xx}]+(uS_3+h)[S,\sigma_3]
$$
$$
\rho u_{tt}=\nu^2_0 u_{xx}+\lambda(S^2_3)_{xx}
$$
the Myrzakulov XXXXVII(M-XXXXVII) [the $ M^{22}_{00}$ - equation]:
$$
2iS_t=[S,S_{xx}]+(uS_3+h)[S,\sigma_3]
$$
$$
\rho u_{tt}=\nu^2_0 u_{xx}+\alpha(u^2)_{xx}+\beta u_{xxxx}+
\lambda (S^2_3)_{xx}
$$
the Myrzakulov XXXXVI(M-XXXXVI) [the $ M^{23}_{00}$ - equation]:
$$
2iS_t=[S,S_{xx}]+(uS_3+h)[S,\sigma_3]
$$
$$
u_t+u_x+\lambda(S^2_3)_x=0
$$
the Myrzakulov XXXXV(M-XXXXV) [the $ M^{24}_{00}$ - equation]:
$$
2iS_t=[S,S_{xx}]+(uS_3+h)[S,\sigma_3]
$$
$$
u_t+u_x+\alpha(u^2)_x+\beta u_{xxx}+\lambda(S^2_3)_x=0
$$

\subsection{The 3-class of spin-phonon systems}

The Myrzakulov XXXXIV(M-XXXXIV) [the $ M^{31}_{00}$ - equation]:
$$
2iS_t=\{(\mu \vec S^2_x - u +m)[S,S_x]\}_x
$$
$$
\rho u _{tt}=\nu^2_0 u_{xx}+\lambda(\vec S^2_x)_{xx}
$$
the Myrzakulov XXXXIII(M-XXXXIII) [the $M^{32}_{00}$ - equation]:
$$
2iS_t=\{(\mu \vec S^2_x - u +m)[S,S_x]\}_x
$$
$$
\rho u _{tt}=\nu^2_0 u_{xx}+\alpha (u^2)_{xx}+\beta u_{xxxx}+ \lambda
(\vec S^2_x)_{xx}
$$
the Myrzakulov XXXXII(M-XXXXII) [the  $M^{33}_{00}$ - equation]:
$$
2iS_t=\{(\mu \vec S^2_x - u +m)[S,S_x]\}_x
$$
$$
u_t+u_x +\lambda (\vec S^2_x)_x = 0
$$
the Myrzakulov XXXXI(M-XXXXI) [the  $M^{34}_{00}$ - equation]:
$$
2iS_t=\{(\mu \vec S^2_x - u +m)[S,S_x]\}_x
$$
$$
u_t+u_x +\alpha(u^2)_x+\beta u_{xxx}+\lambda (\vec S^2_x)_{x} = 0
$$

\subsection{The 4-class of spin-phonon systems}

The Myrzakulov XXXX(M-XXXX) [the  $M^{41}_{00}$ - equation]:
$$
2iS_t=[S,S_{xxxx}]+2\{((1+\mu)\vec S^2_x-u+m)[S,S_x]\}_{x}
$$
$$
\rho u_{tt}=\nu^2_0 u_{xx}+\lambda (\vec S^2_x)_{xx}
$$
the Myrzakulov XXXIX(M-XXXIX) [the  $M^{42}_{00}$ - equation]:
$$
2iS_t=[S,S_{xxxx}]+2\{((1+\mu)\vec S^2_x-u+m)[S,S_x]\}_{x}
$$
$$
\rho u_{tt}=\nu^2_0 u_{xx}+\alpha(u^2)_{xx}+\beta u_{xxxx}+\lambda
(\vec S^2_x)_{xx}
$$
the Myrzakulov XXXVIII(M-XXXVIII) [the  $M^{43}_{00}$ - equation]:
$$
2iS_t=[S,S_{xxxx}]+2\{((1+\mu)\vec S^2_x-u+m)[S,S_x]\}_{x}
$$
$$
u_t + u_x + \lambda (\vec S^2_x)_x = 0
$$
the Myrzakulov XXXVII(M-XXXVII) [the  $M^{44}_{00}$ - equation]:
$$
2iS_t=[S,S_{xxxx}]+2\{((1+\mu)\vec S^2_x-u+m)[S,S_x]\}_{x}
$$
$$
u_t + u_x + \alpha(u^2)_x + \beta u_{xxx}+\lambda (\vec S^2_x)_x = 0
$$

\subsection{The 5-class of spin-phonon systems}

The Myrzakulov XXXVI(M-XXXVI) [the  $M^{51}_{00}$ - equation]:
$$ 2iS_t=[S,S_{xx}]+2iuS_x  $$
$$
\rho u_{tt}=\nu^2_0 u_{xx}+\lambda (f)_{xx}
$$
the Myrzakulov XXXV(M-XXXV) [the  $M^{52}_{00}$ - equation]:
$$ 2iS_t=[S,S_{xx}]+2iuS_x  $$
$$
\rho u_{tt}=\nu^2_0 u_{xx}+\alpha(u^2)_{xx}+\beta u_{xxxx}+\lambda
(f)_{xx} \
$$
the Myrzakulov XXXIV(M-XXXIV) [the  $M^{53}_{00}$ - equation]:
$$ 2iS_t=[S,S_{xx}]+2iuS_x  $$
$$
u_t + u_x + \lambda (f)_x = 0
$$
the Myrzakulov XXXIII(M-XXXIII) [the  $M^{54}_{00}$ - equation]:
$$ 2iS_t=[S,S_{xx}]+2iuS_x  $$
$$
u_t + u_x + \alpha(u^2)_x + \beta u_{xxx}+\lambda (f)_x = 0
$$
Here $f = \frac{1}{4}tr(S^{2}_{x}), \quad \lambda =1. $

\end{document}